\documentclass[aps,prx,twocolumn,floats,showpacs]{revtex4-1}
\usepackage{graphicx}
\usepackage{amsmath}
\usepackage{amssymb}
\usepackage{color}
\usepackage{gensymb}

\newcommand{\opls}[0]{$\mathrm{O^+}$}
\newcommand{\cpls}[0]{$\mathrm{C^+}$}
\newcommand{\rPPO}[0]{$(R)$-PPO}
\newcommand{\sPPO}[0]{$(S)$-PPO}
\newcommand{\sinth}[0]{$\langle \sin(\theta_{2\mathrm{D}}) \rangle$}
\newcommand{\costh}[0]{$\langle \cos^2(\theta_{2\mathrm{D}}) \rangle$}
\newcommand{\dsinth}[0]{$\Delta\langle \sin(\theta_{2\mathrm{D}}) \rangle$}
\newcommand{\avcosth}[0]{$\overline{\langle \cos^2(\theta_{2\mathrm{D}}) \rangle}$}

\begin{document}

\title{Controlled enantioselective orientation of chiral molecules with an optical centrifuge}

\author{Alexander~A.~Milner, Jordan~A.~M.~Fordyce, Ian~MacPhail-Bartley, Walter~Wasserman, and V.~ Milner}
\email{vmilner@phas.ubc.ca}

\affiliation{Department of  Physics \& Astronomy, The University of British Columbia, Vancouver, Canada}

\author{Ilia~Tutunnikov, and  Ilya~Sh.~Averbukh}
\email{ilya.averbukh@weizmann.ac.il}

\affiliation{AMOS and Department of Chemical and Biological Physics, The Weizmann Institute of Science, Rehovot, Israel}

\date{\today}

\begin{abstract}
We report on the first experimental demonstration of enantioselective rotational control of chiral molecules with a laser field. In our experiments, two enantiomers of propylene oxide are brought to accelerated unidirectional rotation by means of an optical centrifuge. Using Coulomb explosion imaging, we show that the centrifuged molecules acquire preferential orientation perpendicular to the plane of rotation, and that the direction of this orientation depends on the relative handedness of the enantiomer and the rotating centrifuge field. The observed effect is in agreement with theoretical predictions and is reproduced in numerical simulations of the centrifuge excitation followed by Coulomb explosion of the centrifuged molecules. The demonstrated technique opens new avenues in optical enantioselective control of chiral molecules with a plethora of potential applications in differentiation, separation and purification of chiral mixtures.
    \end{abstract}

\maketitle

%----------Introduction 1: Chirality --------------------------------
Chirality is a ubiquitous natural phenomenon met in various areas of science and technology ranging from physics of fundamental forces and practical aspects of drug design to studies on the origin of life and biomolecular homochirality \cite{UniversalChirality}. Differentiation and separation of enantiomers in a mixture is an important problem, involving measurements of enantiomeric excess, handedness of a given compound, and devising techniques for manipulating chiral molecules \cite{ShapiroBrumer,bib3,bib4,Patterson2013,Patterson2014,bib8,bib18,bib9,bib11,bib12,bib13,bib14,Cameron_2014,AndrewsBradshaw}.

%----------Introduction 2: Existing methods -------------------------
The standard technique of chiral discrimination - photoabsorption circular dichroism - relies on weak magnetic interactions and yields an extremely small chiral response \cite{Cotton1896,barron_2004}. Methods for chiral discrimination based on electric-dipole interactions have been shown to produce much higher chiral signals. Those include various versions of photo-electron circular dichroism \cite{Ritchie1976,Powis2000,Bowering2001,Beaulieu2018,Steinbacher2015,Beaulieu_2016,Goetz2017,NAHON2015322} (including multiphoton and strong-field ionization regimes \cite{Lux2012,Lehmann2013,Lux_2015,Kastner2016,Beaulieu1288} as well as high harmonic generation \cite{Ayuso2018}), Coulomb explosion imaging \cite{bib8,bib18,Christensen2015,Pitzer2018,Fehre2019}, and enantiosensitive microwave spectroscopy \cite{Patterson2013, Patterson2014}. A unified analysis of this new class of chiral measurements without magnetic interactions can be found in Ref.~\citenum{Ordonez2018}.

%----------Introduction 2: New methods ------------------------------
Recently, a new set of methods has been theoretically proposed for detecting molecular chirality by enantioselective orientation of chiral molecules with strong nonresonant laser fields \cite{Yachmenev2016,Gershnabel2018,Tutunnikov2018}. When linearly polarized, such fields can align an ensemble of molecules along their polarization direction, whereas orienting the molecules is impossible due to the symmetry of the field interaction with the induced dipole. Twisting the polarization in a certain plane breaks that symmetry and introduces a preferred spatial direction perpendicular to the plane of twisting. Twisted fields such as an optical centrifuge \cite{Karczmarek1999,Villeneuve2000,Yuan2011,Korobenko2014}, a pair of two femtosecond pulses \cite{Fleischer2009,Kitano2009,Korech2013,Mizusee1400185,Lin2015}, a chiral pulse train \cite{Zhdanovich2011}, and a polarization shaped pulse \cite{Karras2015} have been used in the past for inducing unidirectional rotation of symmetric molecules. In the case of chiral molecules such fields not only cause the unidirectional rotation of the most polarizable molecular axis in the plane of twisting, but also result in an orienting torque about this axis \cite{Gershnabel2018,Tutunnikov2018}. This torque leads to the molecular orientation either parallel or anti-parallel to the laser beam, depending on the relative sense of polarization rotation and enantiomer handedness. In addition to the detection of molecular chirality, the technique offers an enantioselective \textit{control} of the spatial orientation of chiral molecules, with potential applications to separation and manipulation of chiral mixtures.

%----------Introduction 3: Punch line -------------------------------
Here, we report on the first experimental demonstration of enantioselective control of molecular rotation with laser light. We spin the two enantiomers of propylene oxide (PPO, CH$_{3}$CHCH$_{2}$O) in an optical centrifuge, and show that the centrifuge orients one of the PPO's principal molecular axes either along or against the direction of centrifuge propagation. To detect the chirality-dependent rotation-induced orientation of PPO enantiomers, we employ the techniques of Coulomb explosion and velocity map imaging (VMI). The experimental results are in good qualitative agreement with the numerical analysis, which incorporates both the centrifuge spinning and Coulomb explosion.
\begin{figure}[h]
    \includegraphics[width=1\columnwidth]{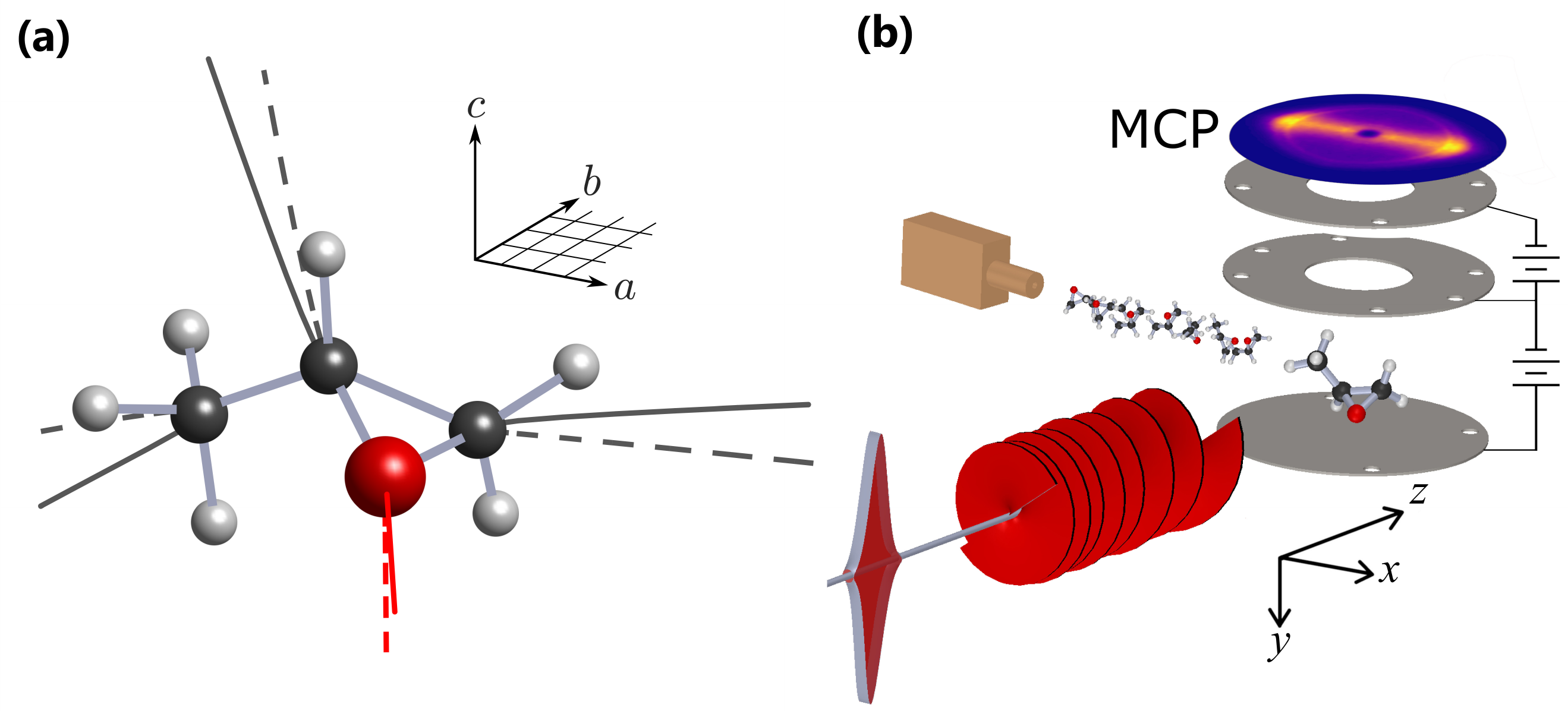}
    \caption{\textbf{(a)} Right-handed enantiomer of propylene oxide, depicted in the frame of its principal axes \textit{a}, \textit{b} and \textit{c}. Red, black and gray spheres represent oxygen, carbon and hydrogen atoms, respectively. Coulomb explosion trajectories from a stationary and a centrifuged molecule are shown with dashed and solid lines (see text for details). \textbf{(b)} Schematic illustration of our experimental geometry. Cold PPO molecules in a seeded molecular jet are spun in an optical centrifuge (long corkscrew shape) and Coulomb exploded with a probe pulse (short diamond shape) between the plates of a conventional velocity map imaging spectrometer, equipped with a multi-channel plate (MCP) detector and a phosphor screen.}
    \label{Fig-Geometry}
\end{figure}

%----------Experimental 1: PPO and jet--------------------------------
Propylene oxide is a relatively simple molecule available for the studies of chirality in the gas phase, and is the only chiral molecule detected in outer space to date \cite{Mcguire2016}. Fig.~\ref{Fig-Geometry}\textbf{(a)} shows the structure of a PPO molecule in the frame of its principal axes, where the moments of inertia are ordered as $I_a<I_b<I_c$. In this work, PPO gas (Sigma Aldrich, $99\%$ purity) was pre-mixed with helium and expanded in vacuum through a 5~$\mu $m diameter nozzle at 20~bar backing pressure. The relative pressure of propylene oxide in the seeded jet was lowered to below $10^{-3}$ with respect to helium for avoiding the formation of PPO dimers, monitored using VMI \cite{Pickering2018}. From the velocity distribution of PPO$^{+}$, the translational temperature of the molecules, which can serve as a rough gauge of their rotational temperature, was estimated as $<5$~K.

%----------Experimental 2: Coulomb explosion and VMI ----------------
Coulomb explosion is a well known method that enables one to extract the information about molecular orientation in the laboratory frame as a function of time \cite{Stapelfeldt1995, Slater2015}. The explosion is initiated by multiple ionization of a molecule with an intense femtosecond probe pulse. This results in a structural destabilization of molecular bonds and separation of positively charged fragments under the action of Coulomb forces. By measuring the velocities of the scattered fragments one can infer the spatial orientation of the molecule at the time of explosion. We use a typical VMI setup in which the molecular jet is intercepted by the probe beam ($\approx50$~fs, $\approx 10^{15}$~W/cm$^{2}$, polarized along $\hat{y}$) between the plates of a time-of-flight spectrometer [Fig.~\ref{Fig-Geometry}\textbf{(b)}]. As the fragment ions accelerate towards, and impinge on, the multi-channel plate detector (MCP), the projection of their velocities on the $xz$ plane is mapped on the plane of the detector. Mass selectivity is provided by gating the MCP at the time of arrival of the fragment of interest.

\begin{figure}[b]
    \includegraphics[width=1\columnwidth]{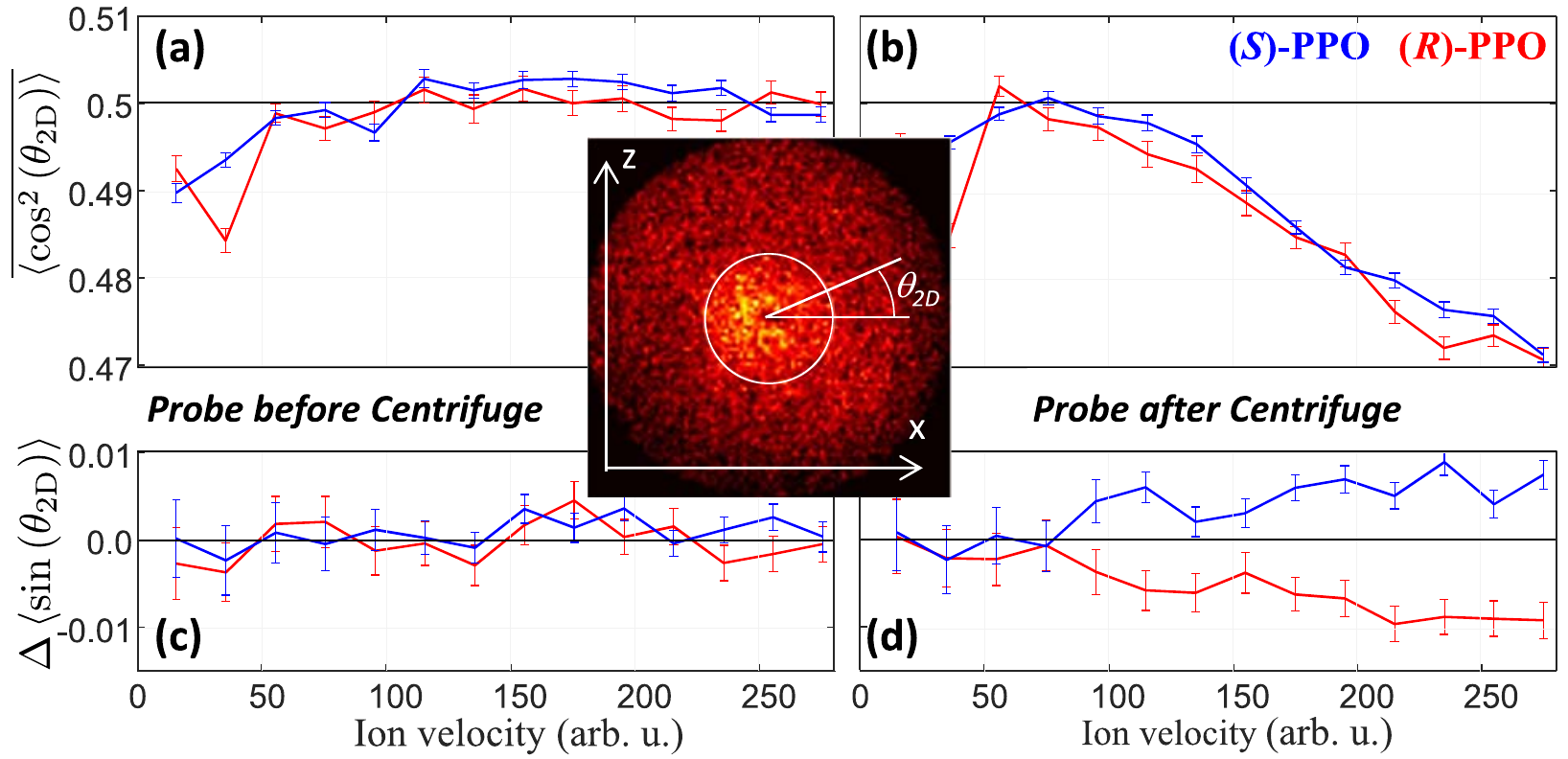}
    \caption{Experimentally measured degree of alignment [upper panels (\textbf{a},\textbf{b})] and orientation [lower panels (\textbf{c},\textbf{d})] of the distribution of \opls{} ions as a function of their velocity. Probe pulses arrived 10~ps \textit{before} [left panels (\textbf{a},\textbf{c})] or \textit{after} [right panels (\textbf{b},\textbf{d})] the arrival of the centrifuge. Blue(red) lines correspond to the left(right)-handed enantiomer. An example of the observed velocity map at the probe delay of $10$~ps is shown in the inset. White circle marks the lower boundary in determining the average values in Fig.~\ref{Fig-Exp2}.}
    \label{Fig-Exp1}
\end{figure}

In this work we are interested in the planar (2D) alignment of PPO's most polarizable \textit{a} axis in the plane of rotation, and the orientation of its \textit{b} axis perpendicular to that plane. The former would confirm the controlled rotation of the molecules in the centrifuge, whereas the latter would show whether this control is sensitive to the handedness of an enantiomer. The two experimental observables, bearing the information about the degree of such alignment and orientation, are conventionally determined as \costh{} and \sinth{}, respectively \cite{Kumarappan2007}. Here $\theta_{2\mathrm{D}}$ is the angle between the ${xz}$ projection of the fragment's velocity $\mathbf{v}$ and the laboratory $x$ axis (see inset in Fig.~\ref{Fig-Exp1}), and $\langle..\rangle$ implies averaging over the molecular ensemble. The value of \costh{} above (below) 0.5 corresponds to the planar alignment (anti-alignment) of $\mathbf{v}$ vectors with respect to the rotation plane. Similarly, positive (negative) values of \sinth{} reflect the orientation of $\mathbf{v}$ along (against) the laboratory $z$ axis. In practice, an average of a few million ion fragments were recorded for each set of experimental conditions, resulting in the precision of $10^{-3}$ in determining both \costh{} and \sinth{}.

%----------Experimental 2: Centrifuge -------------------------------
An optical centrifuge is a laser pulse, whose linear polarization undergoes an accelerated rotation around its propagation direction \cite{Karczmarek1999, Villeneuve2000}. Our setup for producing the centrifuge has been described in an earlier publication \cite{Korobenko14b}. Briefly, we split the spectrum of broadband laser pulses from a Ti:sapphire amplifier (10 mJ, 35 fs, repetition rate 1~KHz, central wavelength 795~nm) in two equal parts using a Fourier pulse shaper. The two beams are first frequency chirped with a chirp rate $\beta $ of equal magnitude and opposite sign ($\beta = \pm 0.3$~ps$^{-2}$). The chirped beams are then combined with a polarizing beam splitter cube and circularly polarized with an opposite sense of circular polarization. Optical interference of these laser fields results in a pulse illustrated in Fig.~\ref{Fig-Geometry}\textbf{(b)}: its polarization vector is rotating in $xy$ plane with an instantaneous angular frequency $\Omega =2\beta  t$.

\begin{figure}[b]
    \includegraphics[width=1\columnwidth]{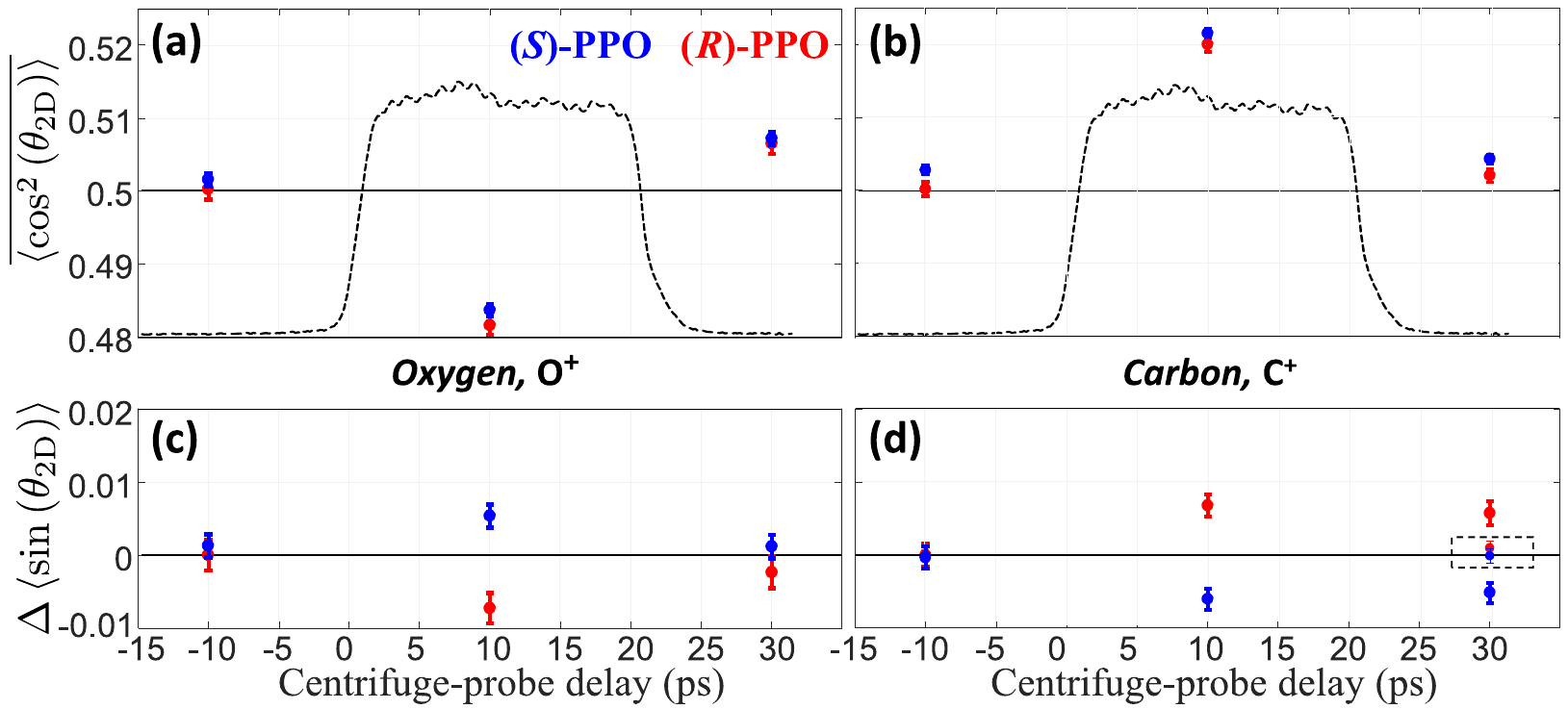}
    \caption{Experimentally measured degree of alignment [upper panels (\textbf{a},\textbf{b})] and orientation [lower panels (\textbf{c},\textbf{d})] in the velocity distribution of \opls{} [left panels (\textbf{a},\textbf{c})] and \cpls{} [right panels (\textbf{b},\textbf{d})] fragments. Blue(red) colored markers correspond to the left(right)-handed enantiomer. Two data points inside the dashed rectangle at 30~ps in panel (\textbf{d}) correspond to the centrifuge with a non-zero initial rotational frequency. The black dashed curves in upper panels show the intensity profile of the centrifuge field in arbitrary units.}
    \label{Fig-Exp2}
\end{figure}

The centrifuge pulse used in this work had an almost rectangular intensity profile with a length of 20~ps (see Fig.~\ref{Fig-Exp2}) and a total energy of 0.5~mJ. The beam is focused on the molecular jet down to a 10~$\mu$m diameter, which results in a peak intensity of $\approx 10^{13}$ W/cm$^{2}$ -- low enough to produce negligible amount of ionized fragments in comparison to the probe-induced rate of Coulomb explosion. The ability to control molecular rotation at such relatively low (compared to alternative methods) laser intensities, thus leaving fragile polyatomic molecules intact, makes the centrifuge particularly well suited for the work on molecular chirality. Our pulse shaper enables full control of both the initial and the terminal frequency of the centrifuge. The former is typically set to zero to ensure the adiabatic spinning, whereas the latter can be varied between 0 and 10~THz.

%----------Exp results 1: O+, COS2(theta) ---------------------------
An example of the recorded ion image of \opls{} fragments is shown in the inset in Fig.~\ref{Fig-Exp1}. When the Coulomb explosion is initiated after the arrival of the centrifuge, a small degree of anisotropy appears in the distribution of fragment velocities. To quantify the degrees of alignment and orientation, we define the following two quantities:
\begin{eqnarray}\label{Eq-Observables}
\overline{\langle \cos^2(\theta_{2\mathrm{D}}) \rangle} &\equiv& 1/2\left[\langle \cos^2(\theta_{2\mathrm{D}}) \rangle_\circlearrowright + \langle \cos^2(\theta_{2\mathrm{D}}) \rangle_\circlearrowleft\right],\nonumber \\
\Delta\langle \sin(\theta_{2\mathrm{D}}) \rangle &\equiv& \langle \sin(\theta_{2\mathrm{D}}) \rangle_\circlearrowright - \langle \sin(\theta_{2\mathrm{D}}) \rangle_\circlearrowleft \nonumber,
\end{eqnarray}
where the indices $\circlearrowright$ and $\circlearrowleft$ correspond to the clockwise and counter-clockwise direction of polarization rotation, as observed along the laser beam propagation.

Prior to the centrifuge arrival [Fig.~\ref{Fig-Exp1}(\textbf{a})] \avcosth{} shows an expected isotropic value of 0.5 at all fragment velocities (aside from the central region suffering from an instrumental artifact). When the molecules are Coulomb exploded 10~ps into their interaction with the centrifuge, the value of \avcosth{} decreases by as much as 0.03, indicating a slight anti-alignment of oxygen trajectories [Fig.~\ref{Fig-Exp1}(\textbf{b})]. Importantly, the effect does not depend on the handedness of the enantiomer (cf. blue and red curves). The observed anti-alignment confirms the rotational excitation of PPO by the centrifuge: as the molecules are spun by the laser field with their most polarizable $a$ axis pulled towards the plane of rotation, the least polarizable $b$ axis, associated with the oxygen atom, tends to stick out perpendicular to that plane.

%----------Exp results 2: O+, SIN(theta) ----------------------------
Similarly to \costh{}, the degree of orientation \sinth{} was found at its isotropic value at negative probe delays, independent of molecular handedness [hence, \dsinth{}$\approx0$ in Fig.~\ref{Fig-Exp1}(\textbf{c})]. However, as the interaction with the centrifuge begins, the orientation factor behaves very differently from the degree of alignment: the two enantiomers show an \textit{opposite sign of orientation} with respect to the molecular handedness and the sense of the polarization rotation. For the left-handed molecule, the probability to find an oxygen fragment in the upper half of the detector $xz$ plane is larger (smaller) than in the lower half if the centrifuge is rotating clockwise (counter-clockwise). This is reflected by \dsinth{}$>0$ for \sPPO{} in Fig.~\ref{Fig-Exp1}(\textbf{d}), where it reaches the values of order of $10^{-2}$. For an opposite enantiomer \dsinth{} is of opposite sign, indicating the central role of chirality in the observed phenomenon.

%----------Exp results 3: O+ vs C+ ----------------------------------
To further support this conclusion, we have collected and analyzed velocity distributions of a \cpls{} fragment. Its behavior is compared to \opls{} in Fig.~\ref{Fig-Exp2}, where we plot the values of \avcosth{} and \dsinth{} averaged over the velocity range between the white circle and the outer boundary on the VMI images in Fig.~\ref{Fig-Exp1}. The average values are shown for three delay times, corresponding to the Coulomb explosion prior to the arrival of the centrifuge (-10~ps), during the centrifuge (10~ps) and after the end of the centrifuge excitation (30~ps). Similarly to the results for oxygen, both the alignment and the orientation of the carbon ion trajectories deviate from their isotropic values once the molecules are exposed to the centrifuge field.

Notably, the distribution of \cpls{} shows an opposite reaction to the centrifuge, as compared with \opls{}. At 10~ps, the alignment factor grows above 0.5 as opposed to dipping below it, and the sign of the orientation factor is reversed. That is, the trajectories of \cpls{} are now aligned, instead of anti-aligned, in the plane of rotation and oriented along the direction of the clockwise centrifuge for the right-handed, instead of left-handed, enantiomers. The reversal of the observed effect reaffirms its rotation-induced nature. Indeed, the two carbon atoms at the opposite ends of a PPO molecule are located close to the molecular $a$ axis and recoil close to the rotation plane upon Coulomb explosion [see Fig.~\ref{Fig-Geometry}(\textbf{a})]. According to this geometric consideration, and inline with our numerical analysis, their combined contribution dominates that of the central C atom, causing the alignment factor averaged over all three carbons to rise above 0.5. On the other hand, the two end C atoms have a much lower effect on the orientation figure, whose sign is therefore dictated by the central carbon. As the latter recoils in the direction approximately opposite to the direction of \opls{}, the orientation sign is flipped.

The enantioselective orientation disappeared beyond our experimental sensitivity in two control cases. In the first one, we repeated the same measurement in a room-temperature background gas of PPO. According to our numerical analysis, hot PPO molecules rotate too quickly to be captured by the centrifuge. We also carried out the same experiment with a centrifuge pulse whose initial rotational frequency was shifted from 0 to 1~THz (with all other parameters kept identical), thus preventing an adiabatic spinning of the molecules. The results of the latter test are shown in Fig.~\ref{Fig-Exp2}(\textbf{d}) within the dashed rectangle. Much lower orientation of ion trajectories in these two cases suggests the rotation-induced mechanism of the observed chiral effect.

%----------Numerics 1: Methods, centrifuge spinning -----------------
To verify these qualitative arguments, we carried out numerical simulations of the experimentally detected fragment ion velocities by modeling both the interaction of a PPO molecule with the field of an optical centrifuge and the following Coulomb explosion of that molecule. The field driven rotational dynamics were described classically and within the rigid rotor approximation by means of Euler equations \cite{LANDAU}. The orientation of the molecule-fixed frame was parametrized using quaternions. The resulting system of coupled Euler equations and quaternion equations of motion was solved numerically. The detailed description of this approach can be found in Ref.~\citenum{Tutunnikov2018}.

%----------Numerics 2: Methods, Coulomb explosion -------------------
We adopted a simplified model of Coulomb explosion which assumes instantaneous conversion of all constituent atoms, whose equilibrium positions do not change during the interaction with the probe pulse, into singly ionized fragments. For finding the asymptotic velocities of each fragment after the Coulomb explosion, we numerically solved the system of coupled Newton's equations. The initial angular velocity of each atom was given by the cross product of its instantaneous angular velocity at the moment of Coulomb explosion and its position vector. To simulate the experimentally observed two-dimensional VMI distribution, we averaged over the thermal ensemble of isotropically distributed molecules, each assigned an angular velocity according to the Boltzmann distribution.

An example of calculated fragment trajectories is shown in Figure~\ref{Fig-Geometry}(\textbf{a}). A static PPO molecule would eject its atoms along the dashed lines. If the molecule was rotating about its $b$ axis with the frequency of 1.9~THz (as if spun by our 20~ps-long centrifuge) the trajectories would be slightly different, as shown by the solid lines. Note that despite the deviations, the velocity of each fragment still carries the naively anticipated information about the spatial orientation of the molecule at the time of explosion. Namely, the oxygen and the middle carbon atoms are recoiling roughly opposite to one another and perpendicular to the molecular $a$ axis, whereas the two end carbons are ejected approximately along that axis. This result supports our interpretation of the observed anisotropy in the experimentally recorded and numerically calculated velocity maps.
\begin{figure}
\begin{centering}
\includegraphics[width=8.4cm]{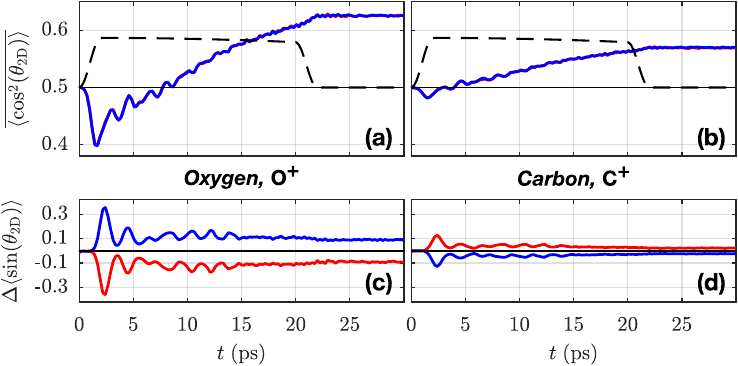}
\par\end{centering}
\caption{Calculated degree of alignment [upper panels (\textbf{a},\textbf{b})] and orientation [lower panels (\textbf{c},\textbf{d})] in the velocity distribution of \opls{} [left panels (\textbf{a},\textbf{c})] and \cpls{} [right panels (\textbf{b},\textbf{d})] fragments for \sPPO{} (blue) and \rPPO{} (red). Horizontal lines mark the respective isotropic values of 0 and $0.5$. Note that the two enantiomers are indistinguishable in terms of their rotation-induced alignment (hence, the fully overlapped upper curves). The black dashed curves in upper panels show the intensity profile of the centrifuge field in arbitrary units.}
\label{fig:observables}
\end{figure}

%----------Numerics 3: Results and comparison -----------------------
Figure~\ref{fig:observables}(\textbf{a}) shows the calculated alignment and orientation factors \avcosth{} and \dsinth{} of oxygen and carbon velocity vectors for both PPO enantiomers. The results for \cpls{} were obtained by averaging over the three alignment/orientation factors of each carbon ion. In these calculations, we used an ensemble of $10^{5}$ molecules and the following simulation parameters (similar to the experimental conditions): peak intensity $I_{0}=2\times 10^{13}\;\mathrm{W/cm^{2}}$, angular acceleration $\beta=\pm0.3\;\mathrm{ps^{-2}}$ and the intensity envelope closely approximating the experimental profile. The molecular parameters for PPO were taken from the NIST database (method B3LYP/cc-pVTZ). The rotational temperature of the initial ensemble was set to $T=10$~K.

The numerical results are in good qualitative agreement with experimental observations. For the alignment of \opls{} velocities, the crossover from below to above the isotropic value of 0.5 is well reproduced [cf. Fig.~\ref{Fig-Exp2}(\textbf{a})]. The average alignment factor for $\mathrm{C}^{+}$ rises above the isotropic value much faster, explaining why its dipping below 0.5 was not observed experimentally. Moreover, the predicted rotation-induced enantioselective molecular orientation \cite{Yachmenev2016, Gershnabel2018, Tutunnikov2018} is indeed imprinted in the orientation of the fragment velocities, thus appearing in both the calculated and measured velocity distributions. Most importantly and similar to the experiment, the sign of the calculated \dsinth{} for the \cpls{} fragment is opposite to that of \opls{}, confirming the orientation of the molecular $b$ axis along or against the propagation direction of the centrifuge.

The predicted magnitude of both the alignment and orientation factors are greater by about an order of magnitude than the experimentally observed. We attribute this disagreement to a number of reasons. In the calculations, we used an oversimplified model of Coulomb explosion which neglects molecular vibration, charge migration as well as the production of polyatomic and multiply ionized fragments. These effects could introduce severe distortions of the fragment trajectories, smearing out the chirality related anisotropy. The limited velocity range of our VMI detector suppresses the effect further by capturing only lower-energy trajectories, whereas its magnitude seems to be growing with energy, as shown in Fig.~\ref{Fig-Exp1}(\textbf{b}). Finally, comparable sizes of our centrifuge and probe beams imply significant contribution from non-rotating molecules, which also decreases the ensemble averaged orientation.

%----------Conclusions ----------------------------------------------
In summary, we report on the enantioselective orientation of chiral molecules by an optical centrifuge. To the best of our knowledge, this is the first experimental demonstration of chiral selectivity in controlling molecular rotation with light. The effect is of classical origin and follows the general scenario of orienting asymmetric molecules by laser fields with twisted polarization \cite{Gershnabel2018,Tutunnikov2018}. When the centrifuge captures and spins the most polarizable molecular axis, the latter lags behind the field polarization vector. This results in an orienting torque due to the nonzero off-diagonal elements of the polarizability tensor, whose sign depends on the handedness of the enantiomer.

Our work demonstrates the potential of an optical centrifuge as a powerful tool for manipulating chiral molecules. Enantioselective orientation, reported here and predicted to last long after the end of the centrifuge pulse \cite{Tutunnikov2018, Tutunnikov2019}, may be used for chiral separation by means of external inhomogeneous fields. Centrifuge spinning is also expected to induce chirality in achiral molecules \cite{Owens2018}, offering new opportunities for studying this fundamental natural phenomenon.

%------Acknowledgement ----------------------------------------------
This work was carried out under the auspices of the Canadian Center for Chirality Research on Origins and Separation (CHIROS), and was partially supported by the Israel Science Foundation (Grant No. 746/15). I.A. and I.T. thank Prof. Paul Brumer for insightful discussions. I.A. acknowledges support as the Patricia Elman Bildner Professorial Chair, and thanks the UBC Department of Physics \& Astronomy for hospitality extended to him during his sabbatical stay. This research was made possible in part by the historic generosity of the Harold Perlman Family.

%\bibliography{Chirality}


\end{document}